\documentclass[12pt]{iopart}
\usepackage{iopams}
\usepackage{graphicx}

\begin{document}

\title{$N$-soliton solutions and perturbation theory for
 the derivative nonlinear Scr\"{o}dinger
equation with nonvanishing boundary conditions}

\author{V M Lashkin}

\address{Institute for Nuclear
Research, Pr. Nauki 47, Kiev 03680, Ukraine}
\ead{vlashkin@kinr.kiev.ua}
\begin{abstract}
We present a simple approach for finding $N$-soliton solution and
the corresponding Jost solutions of the derivative nonlinear
Scr\"{o}dinger equation with nonvanishing boundary conditions.
Soliton perturbation theory based on the inverse scattering
transform method is developed. As an application of the present
theory we consider the action of the diffusive-type perturbation
on a single bright/dark soliton.
\end{abstract}


\section{Introduction}
The derivative nonlinear Schr\"{o}dinger equation (DNLSE)
($\alpha=\pm 1$)
\begin{equation}
\label{main}
i\partial_{t}u+\partial_{x}^{2}u+i\alpha\partial_{x}(|u|^{2}u) =0
\end{equation}
has many physical applications, and, probably, the most important
are in plasma physics and in nonlinear optic. First, equation
(\ref{main}) describes modulated small-amplitude nonlinear
Alfv\'{e}n waves in a low-$\beta$ (the ratio of kinetic to
magnetic pressure) plasma, propagating parallel
\cite{Rogister,Molhus1,Mio} or at a small angle
\cite{Mol86,Molhus2,Kennel} to the ambient magnetic field. The
DNLS equation also describes large-amplitude magnetohydrodynamic
waves in a high-$\beta$ plasma, propagating at an arbitrary angle
to the ambient magnetic field \cite{Ruderman}. In these cases $u$
denotes the transverse magnetic field perturbation normalized by
the ambient magnetic field, where $t$ and $x$ are normalized time
and space coordinates, respectively. Second, the DNLSE is related
to the modified nonlinear Schr\"{o}dinger equation (MNLSE)
\begin{equation}
\label{MNLS}
i\partial_{\tau}\psi+\frac{\sigma}{2}\partial_{\xi}^{2}\psi+is\partial_{\xi}(|\psi|^{2}\psi)
+|\psi|^{2}\psi=0
\end{equation}
by a simple gaugelike transformation \cite{Wadati}
\begin{equation}
\label{gauge}
\psi(\xi,\tau)=u(x,t)\exp\left\{\frac{i}{4s^{4}}t+\frac{i\sigma}{2s^{2}}x\right\},
\end{equation}
where $\sigma=\pm 1$ corresponds to the abnormal (normal) group
velocity dispersion (GVD) region, $\xi=(\sigma
x)/(2s)+t/(2s^{3})$, $\tau=(\sigma t)/(2s^{2})$. In turn, the
MNLSE describes the propagation of ultrashort femtosecond
nonlinear pulses in optical fibers, when the spectral width of the
pulses becomes comparable with the carrier frequency, and, in
addition to the usual Kerr nonlinearity, the effect of
self-steepening of the pulse should be taken into account. In this
case, $u$ is the normalized slowly varying amplitude of the
complex field envelope, $t$ is the normalized propagation distance
along the fiber, and $x$ is the normalized time measured in a
frame of reference moving with the pulse at the group velocity.

Equation (\ref{main}) is completed by the boundary conditions:
vanishing ($u\rightarrow 0$ as $|x|\rightarrow \infty$) or
nonvanishing ($|u|\rightarrow \rho=\mathrm{const}$ as
$|x|\rightarrow \infty$) at infinity. In both cases the DNLSE is
integrable by the inverse scattering transform (IST)
\cite{Kaup,Kawata1,Kawata2,Kawata3}, and admits $N$-soliton
solutions \cite{Steudel}.

The nonvanishing boundary conditions (NVBC) are important in
physical applications. For example, in space plasma physics the
vanishing boundary conditions (VBC) are relevant only for the case
of propagation of Alfv\'{e}n waves strictly parallel to the
ambient magnetic field. In nonlinear optics the NVBC can support
propagation of dark solitons in both normal and abnormal GVD
regions \cite{Chen1}. Unlike the nonlinear Schr\"{o}dinger
equation or the DNLSE with VBC, the DNLSE with NVBC admits
simultaneous generation of breathers (solitons with internal
oscillations) and one-parametric (nonoscillating) bright and/or
dark solitons \cite{Lashkin1}.

The IST formalism for the DNLSE with NVBC is much more complicated
from the one for VBC. Analytical properties of the Jost solutions
in this case are formulated on the Riemann sheets of the spectral
parameter \cite{Kawata1} and the corresponding direct and inverse
scattering problems are rather involved. Recently, Chen and Lam
\cite{Chen1} developed the IST for the DNLSE with NVBC by
introducing an affine parameter to avoid constructing the Riemann
sheets. Both approaches, however, encounter a difficulty when
finding exact explicit $N$-soliton solutions. The reason is that
the resulting solution $u$ containes the phase factor
$\exp(i\eta^{+})$, where $\eta^{+}$ is some definite integral from
$|u|^{2}$. Thus, the solution is written in an implicit form and
only modulus of the solution can be obtained in that way. Though
for simple one-parametric soliton solutions the phase $\eta^{+}$
can be calculated by direct integration, this procedure is
obviously impracticable for $N$-soliton solutions. Instead, tricks
leading to the explicit expression for $\eta^{+}$ were used in
some particular cases: for the two-parametric one-soliton breather
solution \cite{Chen1}, and for the $N$-soliton with purely
imaginary discrete spectral parameters (i. e. for pure bright
and/or dark solitons) \cite{Chen3,Chen4}. Another approach based
on Darboux/B\"{a}clund transformations was developed by Steudel
\cite{Steudel}. Apparently, Steudel was the first to obtain exact
$N$-soliton solutions with explicitly calculated phases for the
DNLSE with NVBC.

From the practical point of view, the completely integrable DNLSE
(\ref{main}) is an idealized model. In many physical applications,
additional terms are often present in the DNLSE. They can include
effects of the third-order linear dispersion, dissipation,
influence of external forces, etc. . These terms violate the
integrability, but being small in many important practical cases,
they can be taken into account by perturbation theory. The most
powerful perturbative technique, which fully uses the natural
separation of the discrete and continuous (i.e., solitonic and
radiative) degrees of freedom of the unperturbed DNLSE, is based
on the IST. While the IST-based perturbation theory for the DNLSE
with VBC was developed long ago \cite{Wyller}, the analogous
theory for nonvanishing boundary conditions was absent.

The aim of this paper is twofold. First, we present a relatively
simple approach for finding exact explicit (i. e. with the phase)
$N+M$-soliton ($N$ breathers and $M$ "usual" bright and/or dark
solitons in asymptotics) solutions of the DNLSE with NVBC and show
that these solutions can be obtained without determining the phase
factor $\exp(i\eta^{+})$. Thus, exact exotic solutions,
describing, for instance, collisions between breathers, as well as
collisions between pure bright/dark solitons and breathers can be
written. Simultaneously, unlike the purely algebraic approach
\cite{Steudel} based on Darboux transformation, the corresponding
Jost solutions can also be obtained. A second aim is to present
perturbation theory for solitons of the DNLSE with NVBC. We derive
evolution equations for the scattering data (both solitonic and
continuous) in the presence of perturbations. As an application of
the present theory we consider the action of the diffusive-type
perturbation on a single bright/dark soliton.

Without loss of generality, we will consider the NVBC in the form
\begin{equation}
\label{bound} u\rightarrow \rho\exp(\pm 2i\theta),\qquad \mbox{as}
\qquad x\rightarrow \pm\infty.
\end{equation}
We also put $\alpha=1$, since the case $\alpha=-1$ can be obtained
from the former by a transformation $x\rightarrow -x$.

The paper is organized as follows. In section 2, we review a
theory of the scattering transform for the DNLSE with NVBC. In
section 3, we present the procedure to construct the general
explicit $N+M$-soliton solution. Integrals of motion are obtained
in section 4. The perturbation theory and its application are
considered in sections 5 and 6, respectively. The conclusion is
made in section 7.

\section{Inverse scattering transform for the DNLSE with NVBC}
\label{sec1}
 In this section we present the theory of the scattering
transform for the DNLSE with NVBC, following \cite{Chen1} with
some modifications and specifications. Equation (\ref{main}) can
be written as the compatibility condition
\begin{equation}
\label{compatib}
\partial_{t}U-\partial_{x}V+[U,V]=0,
\end{equation}
of two linear matrix equations (Kaup-Newell pair) \cite{Kaup}:
\begin{eqnarray}
\label{spec1}
\partial_{x}M(x,t,\lambda)=UM(x,t,\lambda),\\
\label{spec2}
\partial_{t}M(x,t,\lambda)=VM(x,t,\lambda),
\end{eqnarray}
where $\lambda$ is a spectral parameter, and
\begin{eqnarray}
U=-i\lambda^{2}\sigma_{3}+\lambda Q,\qquad \mbox{with } Q=
\left(\begin{array}{cc} 0 & u \\
-u^{\ast} & 0
\end{array}\right),
\\
V=-2i\lambda^{4}\sigma_{3}+2\lambda^{3}Q-i\lambda^{2}Q^{2}\sigma_{3}
+\lambda Q^{3}-i\lambda Q_{z}\sigma_{3}
\end{eqnarray}
Consider the linear problem (\ref{spec1}) for some fixed $t$. In
terms of the matrix $U$ boundary conditions (\ref{bound}) can be
rewritten as
$\lim\limits_{x\rightarrow\pm\infty}U(x,\lambda)=U^{\pm}(\lambda)$,
where
\begin{equation}
U^{\pm}=\left(\begin{array}{cc} -i\lambda^{2} & \rho\lambda e^{\pm 2i\theta} \\
-\rho\lambda e^{\mp 2i\theta} & i\lambda^{2} \end{array}\right).
\end{equation}
Asymptotic solutions of (\ref{spec1}) $E^{\pm}$ satisfy
\begin{equation}
\label{dEdx}
\partial_{x}E^{\pm}=U^{\pm}E^{\pm}.
\end{equation}
The double-valued function
$K(\lambda)=\lambda\sqrt{\lambda^{2}+\rho^{2}}$ appears in the
matrices $E^{\pm}$, and the analytical properties of solutions of
Eq. (\ref{spec1}) are formulated on the Riemann surface determined
by the function $K(\lambda)$.  The Riemann surface $\mathcal{S}$
consists of two sheets $\mathcal{S}^{+}$ and $\mathcal{S}^{-}$ of
the complex $\lambda$--plane with branch cuts on the image axis
from $-\infty$ to $-i\rho$ and from $i\rho$ to $\infty$. It is
convenient to introduce an affine parameter $\zeta$ by a change of
variable \cite{Chen1}
\begin{equation}
\label{change}
\lambda(\zeta)=\frac{1}{2}\left(\zeta-\frac{\rho^{2}}{\zeta}\right).
\end{equation}
This transformation  maps the sheets $\mathcal{S}^{\pm}$ onto
$\mathrm{Im}\, \zeta>0$ and $\mathrm{Im}\,\zeta<0$ respectively
and the real axis on the complex $\lambda$-plane into the real
axis on the $\zeta$-plane.
 Under this,
\begin{equation}
\label{E}
E^{\pm}(x,\zeta)=e^{\pm i\theta\sigma_{3}}\left(\begin{array}{cc} 1 & -i\rho/\zeta \\
-i\rho/\zeta & 1
\end{array}\right)e^{-ik(\zeta)\sigma_{3}x}
\end{equation}
where the single-valued function $k(\zeta)$ is determined by
\begin{equation}
k(\zeta)=\frac{1}{2}\left(\zeta+\frac{\rho^{2}}{\zeta}\right)\lambda(\zeta)
\end{equation}
For $\mathrm{Im}\, \zeta=0$ denote by $M^{\,\pm}(x,\zeta)$ the
$2\times2$ matrix Jost solutions of (\ref{spec1}), satisfying the
boundary conditions
\begin{equation}
\label{bou} M^{\pm}\rightarrow E^{\pm}(x,\zeta), \qquad \mbox{as
}x\rightarrow\pm \infty.
\end{equation}
The corresponding integral equation can be obtained from
(\ref{spec1}) and (\ref{bou})
\begin{equation}
  M^{\pm}(x,\lambda)=E^{\pm}(x,\lambda)\mp
i\lambda\int_{x}^{\pm\infty}E^{\pm}(x-y,\lambda)Q(y)
M^{\pm}(y,\lambda)\,dy. \label{integ}
\end{equation}
The matrix Jost solutions $M^{\,\pm}(x,\zeta)$ can be represented
in the form $M^{-}=(\varphi,\bar{\varphi})$ and
$M^{+}=(\bar{\psi},\psi)$, where $\varphi$ and $\psi$ are
independent vector columns. The monodromy matrix $S(\zeta)$
relates the two fundamental solutions $M^{-}$ and $M^{+}$:
\begin{equation}
\label{S} M^{-}(x,\zeta)=M^{+}(x,\zeta)S(\zeta).
\end{equation}
The Jost coefficients are defined by
\begin{eqnarray}
\varphi=a\bar{\psi}+b\psi,
\label{co1}\\
\bar{\varphi}=-\bar{a}\psi+\bar{b}\bar{\psi},
 \label{co2}
\end{eqnarray}
so that the monodromy matrix is
\begin{equation}
S(\zeta)= \left(\begin{array}{cc}
 a(\zeta)  & -\bar{b}(\zeta) \\
 b(\zeta) & \bar{a}(\zeta)
\end{array}\right),
\end{equation}
where $a\bar{a}+b\bar{b}=1$. Matrices $M^{\pm}$ and $S$ have the
parity symmetry properties
\begin{equation}
\label{parity} S(\zeta)=\sigma_{3}S(-\zeta)\sigma_{3},\qquad
M^{\pm}(\zeta)=\sigma_{3}M^{\pm}(-\zeta)\sigma_{3},
\end{equation}
and the conjugation symmetry properties
\begin{equation}
\label{conjug}
S(\zeta)=\sigma_{2}S^{\ast}(\zeta^{\ast})\sigma_{2},\qquad
M^{\pm}(\zeta)=\sigma_{2}M^{\pm\, \ast}(\zeta^{\ast})\sigma_{2},
\end{equation}
where $\sigma_{2}$ and $\sigma_{3}$ are Pauli matrices, so that
$|a|^{2}+|b|^{2}=1$. In addition, since the scattering problem
(\ref{spec1}) possesses symmetry with respect to the inversion
$\zeta\rightarrow \rho^{2}/\zeta$, the important involution
properties are valid:
\begin{eqnarray}
\label{inv1}
M^{\pm}(x,\rho^{2}/\zeta)=(\zeta/\rho)\sigma_{3}M^{\pm}(x,\zeta)\sigma_{2},\\
\label{inv2}
S(\rho^{2}/\zeta)=\sigma_{2}S(\zeta)\sigma_{2}=S^{\ast}(\zeta^{\ast})
\end{eqnarray}
 It follows from (\ref{S}) that
\begin{equation}
\label{det1} a(\zeta)=\Delta^{-1}(\zeta) \det(\varphi,\psi),
\end{equation}
where we have introduced the notation
\begin{equation}
\Delta(\zeta)\equiv\det\mathbf{M}^{\pm}=1+\rho^{2}/\zeta^{2}.
\end{equation}
 Columns $\varphi(x,\zeta)$ and $\psi(x,\zeta)$ turn
out to be analytically continuable to $\mathrm{Im}\,k(\zeta)>0$
(i. e. to the first and the third quadrants of the complex
$\zeta$-plane), while $\bar{\varphi}$ and $\bar{\psi}$ are
analytically continuable to $\mathrm{Im}\,k(\zeta)<0$ (i. e. to
the second and the fourth quadrants) \cite{Kawata1, Chen1}. Then,
the coefficient $a(\zeta)$ is analytically continuable to $
\mathrm{Im}\,k(\zeta)>0$. The analytic function $a(\zeta)$ may
have zeros $\zeta_{1},\ldots,\zeta_{N}$ in the region of its
analyticity $ \mathrm{Im}\,k(\zeta)>0$. Equation (\ref{det1}) then
shows that the columns $\psi$ and $\varphi$ are linearly dependent
and there exist complex numbers $b_{1},\ldots,b_{N}$ such that
\begin{equation}
\label{gamma} \varphi(x,\zeta_{j})=b_{j}\psi(x,\zeta_{j}),
\end{equation}
and, similarly
\begin{equation}
\label{gamma1}
\bar{\varphi}(x,\zeta_{j}^{\ast})=-b_{j}^{\ast}\bar{\psi}(x,\zeta_{j}^{\ast}).
\end{equation}
The standard analysis of (\ref{integ}) yields the asymptotics at
$|\zeta|\rightarrow\infty$
\begin{eqnarray}
\label{as1} \psi(x,\zeta)e^{-ik(\zeta)x}\rightarrow { -iu/\zeta
\choose 1
}e^{i(\eta^{+}-\theta)}+O(1/|\zeta|^{2}),\\
\varphi(x,\zeta)e^{ik(\zeta)x}\rightarrow{1\choose
-iu^{\ast}/\zeta}e^{i(\eta^{-}-\theta)}+O(1/|\zeta|^{2}),
\end{eqnarray}
where
\begin{equation}
\eta^{\pm}=\pm\frac{1}{2}\int_{x}^{\pm\infty}(\rho^{2}-|u|^{2})\,dx.
\end{equation}
As $|\zeta|\rightarrow 0$, we have
\begin{eqnarray}
\label{as2} \psi(x,\zeta)e^{-ik(\zeta)x}\rightarrow { -i\rho/\zeta
\choose u^{\ast}/\rho
}e^{-i(\eta^{+}-\theta)}+O(1),\\
\varphi(x,\zeta)e^{ik(\zeta)x}\rightarrow{u/\rho \choose
-i\rho/\zeta}e^{-i(\eta^{-}-\theta)}+O(1),
\end{eqnarray}
It then follows from (\ref{det1}) that asymptotics of $a(\zeta)$
are
\begin{eqnarray}
\label{ass1} a(\zeta)\rightarrow
\exp(i\eta-2i\theta), \qquad \mbox{as } |\zeta|\rightarrow\infty \\
\label{ass2} a(\zeta)\rightarrow \exp(-i\eta+2i\theta),\qquad
\mbox{as } |\zeta|\rightarrow 0,
\end{eqnarray}
where
\begin{equation}
\eta=\eta^{+}+\eta^{-}=\frac{1}{2}\int_{-\infty}^{\infty}(\rho^{2}-|u|^{2})\,dx.
\end{equation}
Zeros of $a(\zeta)$ in the region of its analiticity (i. e. to the
first and the third quadrants of the complex $\zeta$-plane) are
not independent due to the symmetry properties (\ref{parity}),
(\ref{conjug}) and (\ref{inv2}) \cite{Chen1}. If $\zeta_{j}$ is a
simple zero of $a(\zeta)$ in the first quadrant, outside the
$\rho$ circle, then $-\zeta_{j}$ (in the third quadrant),
$\rho^{2}/\zeta_{j}^{\ast}$ (in the first quadrant and inside the
$\rho$ circle) and $-\rho^{2}/\zeta_{j}^{\ast}$ (in the third
quadrant and inside the $\rho$ circle) are also simple zeros of
$a(\zeta)$. There are only two zeros for each $j$ if $\zeta_{j}$
lies on the $\rho$ circle: $\zeta_{j}$ and $-\zeta_{j}$. Thus, one
can consider zeros $\zeta_{j}$ lying only in the first quadrant
outside and/or on the $\rho$ circle. These zeros can be
parametrized as $\zeta_{j}=\rho\exp(\gamma_{j}+i\beta_{j})$, where
$\gamma_{j}\geqslant 0$ and $0<\beta_{j}<\pi/2$. In what follows
we assume that in the first quadrant $M$ zeros lie on the $\rho$
circle and $N$ zeros lie outside the $\rho$ circle so that
$j=1\dots M+N$. Using the asymptotics (\ref{ass1}), (\ref{ass2})
and standard methods of the Hilbert transform theory
\cite{Faddeev} in conjunction with the properties (\ref{parity}),
(\ref{conjug}) and (\ref{inv2}), one can express the coefficient
$a(\zeta)$ in terms of its zeros $\zeta_{j}$ in the first quadrant
outside and/or the $\rho$ circle, and the values of $|b(\zeta)|$
on the contour $\Gamma=[0,-\infty]\cup [0,\infty]\cup
[i\infty,0]\cup [-i\infty,0]$
\begin{eqnarray}
 a(\zeta)=e^{i(\eta-2\theta)}\prod_{j=1}^{N}\frac{(\zeta^{2}-\zeta_{j}^{2})}
{(\zeta^{2}-\zeta_{j}^{\ast\,2})}\frac{(\zeta^{2}-\rho^{4}/\zeta_{j}^{\ast\,2})}
{(\zeta^{2}-\rho^{4}/\zeta_{j}^{2})}
 \prod_{k=1}^{M}\frac{(\zeta^{2}-\zeta_{k}^{2})}
{(\zeta^{2}-\zeta_{k}^{\ast\,2})} \nonumber \\ \times
\exp\left\{\frac{1}{2\pi
i}\int_{\Gamma}\frac{\ln(1-|b(\mu)|^{2})}{\mu-\zeta}d\mu\right\}.
\label{S11gen}
\end{eqnarray}
Setting $\zeta=0$ in (\ref{S11gen}) and comparing with
(\ref{as2}), we get
\begin{equation}
\label{eta}
\eta=2\theta-2\sum_{k=1}^{M}\arg\zeta_{k}-4\sum_{j=1}^{N}\arg\zeta_{j}
+\frac{1}{4\pi} \int_{\Gamma}\frac{\ln(|a(\mu)|^{2})}{\mu}d\mu.
\end{equation}
The potential in the general case is
\begin{eqnarray}
\fl u(x)=\rho e^{-2i(\eta^{+}-\theta)}-2\rho
e^{-i(\eta^{+}-\theta)}
\left\{\sum_{j=1}^{N}\left[\frac{c_{j}}{\zeta_{j}}\psi_{1}(x,\zeta_{j})e^{ik_{j}x}+
i\frac{c_{j}^{\ast}}{\rho}\psi_{2}^{\ast}(x,\zeta_{j})e^{-ik_{j}^{\ast}x}\right]
\right. \nonumber \\
+\left.\sum_{k=1}^{M}\frac{c_{j}}{\zeta_{j}}\psi_{1}(x,\zeta_{j})e^{ik_{j}x}-
\frac{1}{2\pi
i}\int_{\Gamma}\frac{r(\zeta)\psi_{1}(x,\zeta)}{\zeta}\,e^{ik(\zeta)x}\,d\zeta\right\},
\end{eqnarray}
where $c_{j}=b_{j}/a'_{j}$ with
$a'_{j}=da/d\zeta|_{\zeta=\zeta_{j}}$. For the compatibilty with
the second Lax equation (\ref{spec2}), the Jost solutions obtained
from (\ref{spec1}) should be multiplied by a $t$-dependent factor
$h(\zeta,t)=\exp[-i\Omega(\zeta)t]$, where
$\Omega(\zeta)=[2\lambda^{2}(\zeta)-\rho^{2}]k(\zeta)$
\cite{Chen1}:
\begin{eqnarray}
\bar{\psi}(x,\zeta,t)=h(\zeta,t)\bar{\psi}(x,\zeta), \qquad
\psi(x,\zeta,t)=h^{-1}(\zeta,t)\psi(x,\zeta),\\
\varphi(x,\zeta,t)=h(\zeta,t)\varphi(x,\zeta), \qquad
\bar{\varphi}(x,\zeta,t)=h^{-1}(\zeta,t)\bar{\varphi}(x,\zeta).
\label{time}
\end{eqnarray}
Dynamics of the scattering data turns out to be trivial
\begin{eqnarray}
a(\zeta,t)=0,\\
b(\zeta,t)=b(\zeta,0)\exp[2i\Omega(\zeta)t],\\
b_{j}(t)=b_{j}(0)\exp[2i\Omega(\zeta_{j})t], \label{b}.
\end{eqnarray}

\section{The Jost solutions and the potential in the
reflectionless case}

An important particular case is that of the reflectionless
(solitonic) potentials $u(x)$ when $b(t,\zeta)\equiv 0$ as a
function of $\zeta$ for some fixed $t$. It then follows from
(\ref{S11gen}) and (\ref{eta}) that
\begin{equation}
\label{a}
a(\zeta)=\prod_{k=1}^{M}\frac{\zeta_{k}^{\ast}(\zeta^{2}-\zeta_{k}^{2})}
{\zeta_{k}(\zeta^{2}-\zeta_{k}^{\ast\,2})}\prod_{j=1}^{N}\frac{\zeta_{j}^{\ast\,2}(\zeta^{2}-\zeta_{j}^{2})}
{\zeta_{j}^{2}(\zeta^{2}-\zeta_{j}^{\ast\,2})}\frac{(\zeta^{2}-\rho^{4}/\zeta_{j}^{\ast\,2})}
{(\zeta^{2}-\rho^{4}/\zeta_{j}^{2})} ,
\end{equation}
which extends to $\mathrm{Im\,\zeta^{2}<0}$ as a meromorphic
function. One also sees that $\bar{a}(t,\zeta)=1/a(t,\zeta)$.
 Since $S(t,\zeta)$ is
diagonal in this case, it can be factorized in such a way
$S^{\,-}(\zeta)=S^{\,+}(\zeta)S(\zeta)$, that the Jost solution
matrices $M^{\,\pm}$ is expressed through a common solution matrix
$A(x,\zeta)$
\begin{equation}
\label{MAS} M^{\,\pm}(x,\zeta)=A(x,\zeta)S^{\,\pm}(\zeta),
\end{equation}
where
\begin{equation}
 \label{S+} S^{\,+}=\left(\begin{array}{cc}
S^{\,+}_{11} & 0 \\ 0 & S^{\,+\, \ast}_{11}
\end{array}\right)
\end{equation}
with
\begin{equation}
\label{S11} S^{\,+}_{11}=\prod_{k=1}^{M}\frac{\zeta_{k}}
{(\zeta^{2}-\zeta_{k}^{2})} \prod_{j=1}^{N}\frac{\zeta_{j}}
{\zeta_{j}^{\ast}(\zeta^{2}-\zeta_{j}^{2})(\zeta^{2}-\rho^{4}/\zeta_{j}^{\ast\,2})},
\end{equation}
 and
\begin{equation}
\label{S-} S^{\,-}=\sigma_{1}S^{\,+}\sigma_{1},
\end{equation}
where $\sigma_{1}$ is Pauli matrix. One can see from (\ref{S+})
and (\ref{S-}) that $S^{\,+}(\rho)=S^{\,-}(\rho)$ if $M$ is even,
and $S^{\,+}(\rho)=S^{\,-}(\rho)$ if $M$ is odd. Therefore, from
(\ref{MAS}) we get
\begin{equation}
\label{M} M^{\,+}(x,\rho)=(-1)^{M}M^{\,-}(x,\rho).
\end{equation}
On the other hand, since $\lambda=0$ corresponds to
$\zeta=\pm\rho$, we have from  (\ref{E}) and (\ref{integ})
\begin{equation}
\label{E1}
M^{\pm}(x,\rho)=E^{\pm}(x,\rho)=e^{\pm i\theta\sigma_{3}}\left(\begin{array}{cc} 1 & -i \\
-i & 1
\end{array}\right).
\end{equation}
It then immediately follows from (\ref{M}) and (\ref{E1}) that
$\theta=0\pm \pi n$ if $M$ is even, and $\theta=\pm \pi/2\pm \pi
n$ if $M$ is odd ($n$ is an integer). Thus, we established the
following important fact: the total phase shift $4\theta$ in the
$N$-soliton solution is zero (or an integer times $2\pi$). Note,
that authors of \cite{Molhus2, Chen1} showed that $\theta=0$ for
the particular case $N=1$ considering an explicit, rather
complicated expression for the one-soliton breather solution. From
(\ref{MAS})-(\ref{S11}) and (\ref{E1}) one can also obtain
\begin{equation}
\fl A(x,\rho)=\rho^{2N}(-\sigma_{1})^{M}\left(\begin{array}{cc} 1 & -i \\
-i & 1
\end{array}\right)\prod_{k=1}^{M}(\zeta_{k}^{\ast}-\zeta_{k})
\prod_{j=1}^{N}\frac{[\rho^{2}(\zeta_{j}^{\ast \,
2}+\zeta_{j}^{2})-\rho^{4}-|\zeta_{j}|^{4} ]}{|\zeta_{j}|^{2}},
\label{Ar}
\end{equation}
As follows from (\ref{inv1}), (\ref{gamma}) and (\ref{gamma1}) the
columns of $A(x,\zeta)$ satisfy the relations
 \numparts
\label{system}
\begin{eqnarray}
\label{ur1}
A_{1}(x,\zeta_{j})=b_{j}A_{2}(x,\zeta_{j}),\\
\label{ur2}
A_{2}(x,\zeta_{j}^{\ast})=-b_{j}^{\ast}A_{1}(x,\zeta_{j}^{\ast}),\\
\label{ur3}
A_{1}(x,\rho^{2}/\zeta_{j}^{\ast})=b_{j}^{\ast}A_{2}(x,\rho^{2}/\zeta_{j}^{\ast}),\\
\label{ur4}
A_{2}(x,\rho^{2}/\zeta_{j})=-b_{j}A_{1}(x,\rho^{2}/\zeta_{j}),
\end{eqnarray}
\endnumparts
 for all $j=1,\ldots,M+N$. For $M$ zeros lying on the
$\rho$ circle, equations (\ref{ur1})-(\ref{ur4}) become
\begin{eqnarray}
\label{u5}
A_{1}(x,\zeta_{j})=b_{j}A_{2}(x,\zeta_{j}),\\
\label{u6}
A_{2}(x,\zeta_{j}^{\ast})=-b_{j}A_{1}(x,\zeta_{j}^{\ast}),
\end{eqnarray}
where the coefficients $b_{j}$ are real. One can see from
(\ref{as2}) and (\ref{MAS}) that $A(\zeta)$ is analytical in the
whole $\zeta$ plane, except for the point $\zeta=0$, where
off-diagonal elements of $A(\zeta)\exp[ik(\zeta)x\sigma_{3}]$ have
simple pole. Thus, the matrix $\zeta
A(\zeta)\exp[ik(\zeta)x\sigma_{3}]$ is analytical in the whole
$\zeta$ plane. It then follows from (\ref{as1}) and (\ref{MAS})
that diagonal and off-diagonal elements of the matrix $\zeta
A(\zeta)\exp[ik(\zeta)x\sigma_{3}]$ are polynomials in $\zeta$ of
degrees $4N+2M+1$ and $4N+2M$, respectively. In addition, from
(\ref{parity}) and (\ref{MAS}) one sees that the diagonal elements
of $A$ are even in $\zeta$, while the off-diagonal ones are odd.
This means that we can write
\begin{equation}
\label{Ae} \fl  A(x,\zeta)e^{ik(\zeta)x\sigma_{3}}=
\left(\begin{array}{cc}  A_{11}^{(0)} & A_{12}^{(0)}\zeta^{-1} \\
A_{21}^{(0)}\zeta^{-1} & A_{22}^{(0)}\end{array}\right)
+\sum_{p=1}^{L}\zeta^{2p-1}
\left(\begin{array}{cc} \zeta A_{11}^{(p)} & A_{12}^{(p)}\\
A_{21}^{(p)} & \zeta A_{22}^{(p)}\end{array}\right), \label{A}
\end{equation}
where $L=2N+M$, and $A_{mn}^{(p)}$ are some unknown functions of
$x$. Setting $\zeta=\rho$ in (\ref{A}) and comparing with
(\ref{Ar}), one can get
\begin{eqnarray}
\fl \left(\begin{array}{cc}  A_{11}^{(0)} & A_{12}^{(0)}\rho^{-1} \\
A_{21}^{(0)}\rho^{-1} & A_{22}^{(0)}\end{array}\right)
+\sum_{p=1}^{L}\rho^{2p-1}
\left(\begin{array}{cc} \rho A_{11}^{(p)} & A_{12}^{(p)}\\
A_{21}^{(p)} & \rho A_{22}^{(p)}\end{array}\right)
=\rho^{2N}(-\sigma_{1})^{M}\nonumber \\
\times\left(\begin{array}{cc} 1 & -i \\
-i & 1
\end{array}\right)\prod_{k=1}^{M}(\zeta_{k}^{\ast}-\zeta_{k})
\prod_{j=1}^{N}\frac{[\rho^{2}(\zeta_{j}^{\ast \,
2}+\zeta_{j}^{2})-\rho^{4}-|\zeta_{j}|^{4} ]}{|\zeta_{j}|^{2}} .
\label{ur5}
\end{eqnarray}
The unknown coefficients $A_{mn}^{(p)}(x,t)$ with $p=0,\dots ,L$
are determined uniquely from (\ref{system}) and (\ref{ur5}).
Indeed, the first row of (\ref{system}) and (\ref{ur5}) is a
linear algebraic system of $2L+2$ equations in $2L+2$ unknowns,
the coefficients $A_{12}^{(p)}$ and $A_{11}^{(p)}$ with
$p=0,\ldots,L$. Likewise, the second row of (\ref{system}) and
(\ref{ur5}) is the system for determining $A_{21}^{(p)}$ and
$A_{22}^{(p)}$ with $p=0,\ldots,L$.
 By direct substitution
one can check that (\ref{A}) is compatible with (\ref{spec1}) and
(\ref{MAS}) if and only if
\begin{equation}
\label{N-sol}
 u(x,t)=\frac{i\,A_{12}^{(\,L)}(x,t)}{
A_{22}^{(\,L)}(x,t)}.
\end{equation}
This formula reconstructs $u(x,t)$ from the discrete scattering
data $\{\zeta_{j}(t)\}$, $\{b_{j}(t)\}$ in the case when
$b(t,\zeta)\equiv 0$ and it gives $N+M$-soliton solution of
(\ref{main}). An explicit form of the solution can be easily
written in terms of the determinants of corresponding matrices.
Equations (\ref{MAS}) and (\ref{A}) determine the $N+M$-soliton
Jost solutions.

\begin{figure}
\includegraphics[width=6in]{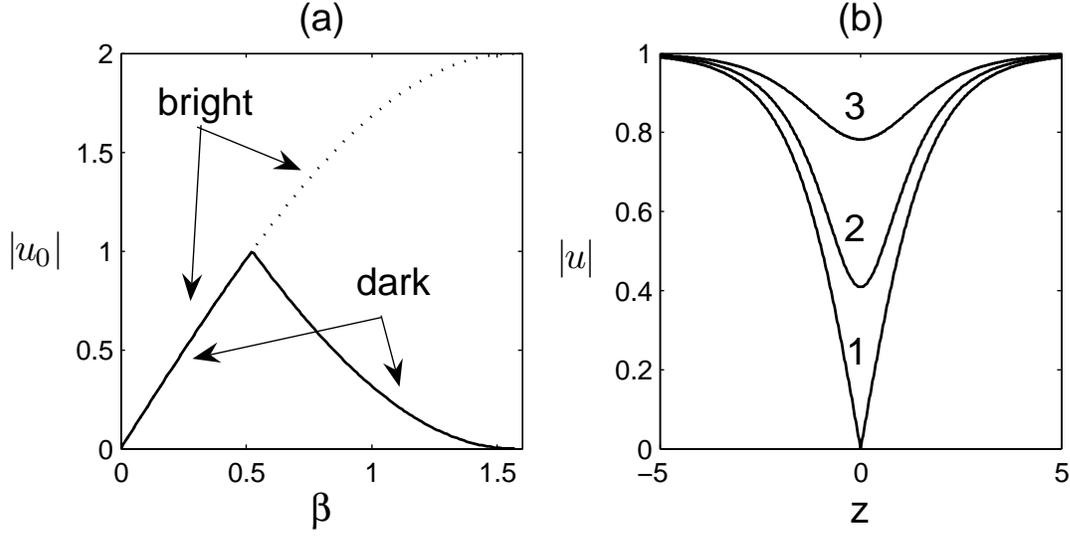}
\caption{\label{fig1} (a) Bright and dark soliton amplitudes
versus the parameter $\beta$ for $\rho=1$. The dotted line and the
solid line up to the bifurcation point correspond to the bright
soliton. The solid line corresponds to the dark soliton. The
bifurcation point is at $\beta_{cr}=\pi/6$ and $|u_{0}|=\rho$. (b)
Dark soliton profiles for different $\beta$ and $\rho=1$. Curves
$1$ ("black" soliton), $2$ and $3$ correspond to
$\beta_{cr}=\pi/6$, $\beta=0.3$, and $\beta=1.1$, respectively. }
\end{figure}

As the first example, let us consider the simplest case when the
function $a(\zeta)$ has one simple zero $\zeta_{1}$ in the first
quadrant of the complex $\zeta$-plane on the $\rho$ circle (i. e.
$M=1$, $N=0$) so that $\zeta_{1}=\rho\exp(i\beta_{1})$ with
$0<\beta_{1}<\pi/2$. Taking into account Eq. (\ref{b}), we have
$b_{1}\exp(2ik_{1}x)=\epsilon\exp(-z)$, where
$z=k_{0}(x-vt-x_{0})$ with
\begin{equation}
\label{kv} k_{0}=\rho^{2}\sin(2\beta_{1}), \qquad
v=2\rho^{2}-\rho^{2}\cos(2\beta_{1}),
\end{equation}
and without loss of generality one can set $\epsilon=\pm 1$.
Determining $A_{12}^{(0)}$ and $A_{11}^{(0)}$ from (\ref{ur5}) and
solving a system of two linear algebraic equations for
$A_{12}^{(1)}$ and $A_{11}^{(1)}$ from (\ref{u5}), (\ref{u6}) we
get
\begin{equation}
A_{12}^{(1)}=e^{-i\beta_{1}}\frac{(e^{3i\beta_{1}}-i\epsilon
e^{-z})}{(e^{i\beta_{1}}+i\epsilon e^{-z})},\qquad
A_{11}^{(1)}=\frac{(i+\epsilon
e^{i\beta_{1}-z})}{\rho(e^{i\beta_{1}}+i\epsilon e^{-z})}
\end{equation}
The one-soliton solution is $u(x,t)=iA_{12}^{(1)}/A_{22}^{(1)}$,
and taking into account the property
$A_{22}^{(j)}=A_{11}^{(j)\,\ast}$, we have
\begin{equation}
\label{usual} u(x,t)=\rho\left[1-\frac{2i
\cos^{2}\beta_{1}}{\epsilon\sinh(z+i\beta_{1}) +i}\right].
\end{equation}
The case $\epsilon=-1(1)$ corresponds to bright (dark) soliton.
The parameters $k_{0}$ and $v$ in (\ref{kv})
 are the soliton inverse width and the soliton
velocity, respectively. In fact, there is only one parameter
$\beta_{1}$ characterizing the soliton, and it is usually called a
one-parametric soliton \cite{Molhus2}. Amplitudes (with respect to
the background) of the bright and dark solitons are
$A_{b}=2\rho\sin\beta_{1}$ and $A_{d}=\rho-\rho
|1-2\sin\beta_{1}|$, respectively. Dependences of the amplitudes
on the parameter $\beta$ are shown in figure \ref{fig1}(a). It is
interesting to note that the dark soliton amplitude is a
nonmonotonic function of $\beta$ and the maximum occurs at
$\beta_{cr}=\pi/6$. The dark soliton profiles $|u(z)|$ for
different $\beta$ and $\rho=1$ are presented in figure
\ref{fig1}(b). The dark soliton with $\beta=\beta_{cr}$ (the curve
$1$ in figure \ref{fig1}(b)) may be called "black" soliton: the
intensity in the center of the soliton falls to zero. The
corresponding one-soliton Jost solutions can easily be obtained
from equations (\ref{time}), (\ref{MAS}) and (\ref{Ae})
\begin{equation}
\bar{\psi}_{1}(x,\zeta,t)=\frac{e^{-ik(\zeta)x}\zeta_{1}}{(\zeta^{2}-\zeta_{1}^{2})}
[2\rho\sin\beta_{1}+(\zeta^{2}-\rho^{2})A_{11}^{(1)}]h(\zeta,t),
\end{equation}
\begin{equation}
\label{psi1}
\psi_{1}(x,\zeta,t)=\frac{e^{ik(\zeta)x}\zeta_{1}^{\ast}}{\zeta(\zeta^{2}-
\zeta_{1}^{\ast\,2})}[2i\rho^{2}\sin\beta_{1}+(\zeta^{2}-\rho^{2})A_{12}^{(1)}]
h^{-1}(\zeta,t),
\end{equation}
\begin{equation}
\varphi_{1}(x,\zeta,t)=\frac{e^{-ik(\zeta)x}\zeta_{1}^{\ast}}{(\zeta^{2}-\zeta_{1}^{\ast\,2})}
[2\rho\sin\beta_{1}+(\zeta^{2}-\rho^{2})A_{11}^{(1)}]h(\zeta,t),
\end{equation}
\begin{equation}
\bar{\varphi}_{1}(x,\zeta,t)=\frac{e^{ik(\zeta)x}\zeta_{1}}{\zeta(\zeta^{2}-
\zeta_{1}^{2})}[2i\rho^{2}\sin\beta_{1}+(\zeta^{2}-\rho^{2})A_{12}^{(1)}]
h^{-1}(\zeta,t).
\end{equation}
The remaining Jost solutions can be found from the symmetry
properties (\ref{parity}), (\ref{conjug}) and (\ref{inv1})
\begin{equation}
\label{psi2} \psi_{2}=\bar{\psi}_{1}^{\ast},\qquad
\bar{\psi}_{2}=-\psi_{1}^{\ast}, \qquad
\bar{\varphi}_{2}=\varphi_{1}^{\ast}, \qquad
\varphi_{2}=-\bar{\varphi}_{1}^{\ast}.
\end{equation}
Next we write out solutions for two more cases: the case when
$a(\zeta)$ has two simple zeros in the first quadrant on the
$\rho$ circle (i. e. $M=2$, $N=0$) so that
\begin{equation}
\label{twosol} \zeta_{1}=\rho\exp(i\beta_{1}), \qquad
\zeta_{2}=\rho\exp(i\beta_{2}),
\end{equation}
and the case when $a(\zeta)$ has one simple zero in the first
quadrant outside the $\rho$ circle (i. e. $M=0$, $N=1$)
\begin{equation}
\label{breath} \zeta_{1}=\rho\exp(\gamma_{1}+i\beta_{1}),\qquad
\gamma_{1}>0.
\end{equation}
The case (\ref{twosol}) corresponds to two-soliton solution for
the one-parametric solitons, while the case (\ref{breath})
corresponds to two-parametric one-soliton solution.
 In both cases we need to solve a system of four
linear algebraic equations. Under this, the corresponding minors
and determinants can be factorized and some parts of them are
cancelled so that the resulting expressions for $u$ can
significantly be simplified. The solutions are of the form
\begin{equation}
\label{N2} u=\rho\frac{BD}{D^{\ast\,2}},
\end{equation}
where for the case (\ref{twosol})
\begin{eqnarray}
\fl B=1-i\epsilon_{1}e^{-3i\beta_{1}-z_{1}}-
i\epsilon_{2}e^{-3i\beta_{2}-z_{2}}-\epsilon_{1}\epsilon_{2}
\frac{\sin^{2}(\beta_{1}-\beta_{2})}{\sin^{2}(\beta_{1}+\beta_{2})}
e^{-3i(\beta_{1}+\beta_{2})}e^{-z_{1}-z_{2}},
\label{col1}\\
\fl D=1-i\epsilon_{1}e^{i\beta_{1}-z_{1}}-
i\epsilon_{2}e^{i\beta_{2}-z_{2}}-\epsilon_{1}\epsilon_{2}
\frac{\sin^{2}(\beta_{1}-\beta_{2})}{\sin^{2}(\beta_{1}+\beta_{2})}
e^{i(\beta_{1}+\beta_{2})}e^{-z_{1}-z_{2}}, \label{col2}
\end{eqnarray}
where $z=k_{0,j}(x-v_{j}t-x_{0,j})$ ($j=1,2$) with
\begin{equation}
\label{kvj} k_{0,j}=\rho^{2}\sin(2\beta_{j}), \qquad
v_{j}=2\rho^{2}-\rho^{2}\cos(2\beta_{j}),
\end{equation}
and, as before, $\epsilon_{j}=-1(1)$ corresponds to bright (dark)
soliton. Equations (\ref{N2}), (\ref{col1}) and (\ref{col2})
describe collisions between bright/dark and bright/dark solitons.

 For the case (\ref{breath}) we get
\begin{eqnarray}
\fl B=\sinh 2\gamma_{1}\cosh(z+2\gamma_{1}+3i\beta_{1}-\ln\sinh 2
\gamma_{1})+ \sin 2\beta_{1}\sinh(3\gamma_{1}-i\varphi),
\label{breath1} \\
\fl D=\sinh 2 \gamma_{1}\cosh(z+2\gamma_{1}-i\beta_{1}-\ln\sinh 2
\gamma_{1})- \sin 2\beta_{1}\sinh(\gamma_{1}+i\varphi),
\label{breath2}
\end{eqnarray}
with
\begin{equation}
\fl z=k_{0}(x-vt-x_{0}), \qquad \varphi=\mu(x-wt)+\varphi_{0},
\end{equation}
\begin{equation}
\fl k_{0}=\rho^{2}\cosh 2\gamma_{1}\sin 2\beta_{1}, \qquad
\mu=\rho^{2}\sinh 2\gamma_{1}\cos 2\beta_{1},
\end{equation}
\begin{equation}
\fl \label{vb} v=2\rho^{2}-\rho^{2}\cos 2\beta_{1}\frac{\cosh
4\gamma_{1}}{\cosh 2\gamma_{1}},\qquad w=2\rho^{2}-\rho^{2}\cosh
2\gamma_{1}\frac{\cos 4\beta_{1}}{\cos 2 \beta_{1}}.
\end{equation}
The two-parametric soliton given by (\ref{N2}), (\ref{breath1})
and (\ref{breath2}) with the parameters $\gamma_{1}$ and
$\beta_{1}$ is actually a breather (oscillating soliton) with
period
\begin{equation}
T=\frac{2\pi}{\rho^{2}\tanh(2\gamma_{1})[\cosh^{2}(2\gamma_{1})+\cos^{2}(2\beta_{1})]},
\end{equation}
and with velocity $v$ given by (\ref{vb}). If
$\gamma_{1}\rightarrow 0$ and $\varphi_{0}\neq n\pi$ ($n$ is an
integer), we have $T\rightarrow\infty$ and the breather reduces to
the one-parametric soliton (bright or dark, depending on
$\varphi_{0}$) given by (\ref{usual}). The found soliton solutions
perfectly coincide with those obtained in \cite{Chen1,Chen3}.

\section{Integrals of motion}
Being completely integrable, the DNLSE with NVBC has an infinite
set of integrals of motion. Eliminating $\varphi_{2}$ from
(\ref{spec1}), and substituting
\begin{equation}
\label{mo1}
\varphi_{1}=\exp\{-i\theta-ik(\zeta)x+i\eta^{-}(x)+q(x,\zeta)\},
\end{equation}
into the  resulting equation for $\varphi_{1}$, we get Riccati
equation for the function $f=i(\rho^{2}-|u|^{2})/2+\partial_{x}q$
\begin{equation}
\label{ric}
\partial_{x}f+(f-ik)^{2}-\frac{u'}{u}(f-ik)+
\lambda^{2}\left(|u|^{2}-i\frac{u'}{u}+\lambda^{2}\right)=0,
\end{equation}
where $u'\equiv \partial_{x}u$. Representing
\begin{equation}
\label{rep}
f(x,\zeta)=\frac{1}{i}\sum_{n=0}^{\infty}\frac{f_{n}(x)}{\zeta^{2n}},
\end{equation}
and substituting (\ref{rep}) into (\ref{ric}), one can
successively determine the coefficients $f_{n}(x)$. The first few
of them are
\begin{eqnarray}
\fl
f_{0}=\frac{1}{2}(|u|^{2}-\rho^{2}),\\
\fl
f_{1}=-iu\partial_{x}u^{\ast}-\frac{1}{2}(|u|^{4}-\rho^{4}),\\
\fl f_{2}=2i(|u|^{2}-\rho^{2})u\partial_{x}u^{\ast}
+i|u|^{2}u^{\ast}\partial_{x}u
-2u\partial_{x}^{2}u^{\ast}+|u|^{4}(|u|^{2}-\rho^{2}).
\end{eqnarray}
From equations (\ref{E}) and (\ref{bou}) we have
\begin{eqnarray}
\label{as3} \varphi_{1}\rightarrow
e^{-i\theta-ikx}, \qquad \mbox{as } x\rightarrow -\infty \\
\label{as4} \varphi_{1}\rightarrow
ae^{i\theta-ikx}-\frac{i\rho}{\zeta}be^{i\theta+ikx},\qquad
\mbox{as } x\rightarrow \infty.
\end{eqnarray}
It then follows from $\eta^{-}(-\infty)=0$ and from equation
(\ref{mo1}) that $q(-\infty,\zeta)=0$. Since
$\eta^{-}(\infty)=\eta$, from equations (\ref{mo1}) and
(\ref{as4}) one can find that $q(\infty,\zeta)=\ln
a(\zeta)+2i\theta-i\eta$ as $x\rightarrow\infty$ and
$|\zeta|\rightarrow \infty$. On the other hand, from the
definition of the function $f(x,\zeta)$ we have
\begin{equation}
q(\infty,\zeta)=\int_{-\infty}^{\infty}f(x,\zeta)\,dx-i\eta.
\end{equation}
Thus, taking into account (\ref{rep}), one obtains
\begin{equation}
\label{con1} \ln
a(\zeta)=-2i\theta-i\sum_{n=0}^{\infty}\frac{I_{n}}{\zeta^{2n}},
\end{equation}
where
\begin{equation}
I_{n}=\int_{-\infty}^{\infty}f_{n}(x)\,dx
\end{equation}
are integrals of motion. As usual, expanding (\ref{S11gen}) in
power series with respect to $1/\zeta$ and using (\ref{con1}), one
can explicitly express the integrals of motion in terms of
discrete (solitonic) and continuous scattering data. In
particular, for $I_{0}$ we get equation (\ref{eta}), and for
$I_{1}$ we have
\begin{eqnarray}
\fl
I_{1}=i\sum_{j=1}^{N}\left[\rho^{4}\left(\frac{1}{\zeta_{j}^{2}}-
\frac{1}{\zeta_{j}^{\ast\,2}}\right)+(\zeta_{j}^{\ast\,2}-\zeta_{j}^{2})\right]+
i\sum_{k=1}^{M}(\zeta_{k}^{\ast\,2}-\zeta_{k}^{2}) \nonumber \\
 - \frac{1}{2\pi}\int_{\Gamma}\mu\ln(1-|b(\mu)|^{2})\,d\mu.
\end{eqnarray}

\section{Perturbation theory}
In the presence of perturbations the DNLSE can be written as
\begin{equation}
\label{mainp}
i\partial_{t}u+\partial_{x}^{2}u+i\partial_{x}(|u|^{2}u)
=p[u,u^{\ast}],
\end{equation}
where the perturbation is represented by the term $p[u,u^{\ast}]$.
Equation (\ref{mainp}) can be cast in the matrix form
\begin{equation}
\label{cast}
\partial_{t}U-\partial_{x}V+[U,V]+P=0,
\end{equation}
where
\begin{equation}
 P=
\left(\begin{array}{cc} 0 & i\lambda p \\
-i\lambda p^{\ast} & 0
\end{array}\right).
\end{equation}
Then, evolution equation for the monodromy matrix $S$ can be
obtained in a way similar to that described in \cite{Lashkin2}. As
a result, we have
\begin{equation}
\label{motionS} \fl  \partial_{t} S(t,\zeta)+i\Omega(\zeta)
[\sigma_{3},S(t,\zeta)]
 =-\int_{-\infty}^{\infty}(M^{+})^{-1}(x,t,\zeta)P
 M^{-}(x,t,\zeta)dx.
\end{equation}
The equations of motion for the coefficients $a(t,\zeta)$ and
$b(t,\zeta)$ are contained in Eq. (\ref{motionS}). Taking into
account that $\det M^{\pm}=1+\rho^{2}/\zeta^{2}$ and equation
(\ref{change}), we have
\begin{equation}
\label{eqa} \frac{\partial a}{\partial
t}=\frac{i\zeta(\rho^{2}-\zeta^{2})}{(\rho^{2}+\zeta^{2})}\int_{-\infty}^{\infty}
(p\psi_{2}\varphi_{2}+p^{\ast}\psi_{1}\varphi_{1})\,dx,
\end{equation}
\begin{equation}\label{eqb}
 \frac{\partial b}{\partial
t}-2i\Omega(\zeta)b=-\frac{i\zeta(\rho^{2}-\zeta^{2})}{(\rho^{2}+\zeta^{2})}
\int_{-\infty}^{\infty}
(p\bar{\psi_{2}}\varphi_{2}+p^{\ast}\bar{\psi_{1}}\varphi_{1})\,dx.
\end{equation}
 The
expression defining the zeros $\zeta_{j}(t)$ of $a(t,\zeta)$ is
$a(t,\zeta_{j}(t))=0$. Differentiating with respect to $t$ gives
\begin{equation}
\label{ta}
\partial_{t}a(t,\zeta_{j}(t))+\frac{\partial\zeta_{j}}{\partial t}\,
a'_{j}=0,
\end{equation}
where $a'_{j}=da/d\zeta|_{\zeta=\zeta_{j}}$. Using (\ref{eqa}) and
(\ref{ta}), one therefore finds
\begin{equation}
\label{zeta} \frac{\partial\zeta_{j}}{\partial
t}=-\frac{i\zeta_{j}(\rho^{2}-\zeta_{j}^{2})}{(\rho^{2}+\zeta_{j}^{2})a'_{j}}
\int_{-\infty}^{\infty}
(p\psi_{2,j}\varphi_{2,j}+p^{\ast}\psi_{1,j}\varphi_{1,j})\,dx,
\end{equation}
or, taking into account (\ref{gamma}),
\begin{equation}
\label{zeta1} \frac{\partial\zeta_{j}}{\partial
t}=-\frac{i\zeta_{j}(\rho^{2}-\zeta_{j}^{2})b_{j}}{(\rho^{2}+\zeta_{j}^{2})a'_{j}}
\int_{-\infty}^{\infty}
(p\psi_{2,j}^{2}+p^{\ast}\psi_{1,j}^{2})\,dx,
\end{equation}
where $\psi_{2,j}$, $\varphi_{2,j}$, $\psi_{1,j}$, and $
\varphi_{1,j}$ are the corresponding Jost solutions evaluated at
$\zeta=\zeta_{j}$. Evolution equation for $b_{j}$ can be obtained
in a way similar to that described in Ref. \cite{Lashkin2}. As a
result, one obtaines
\begin{eqnarray}
 \frac{\partial b_{j}}{\partial
t}-2i\Omega(\zeta_{j})b_{j}=-\frac{i\zeta_{j}(\rho^{2}-\zeta_{j}^{2})}
{(\rho^{2}+\zeta_{j}^{2})a'_{j}} \int_{-\infty}^{\infty}\left\{
p\,\varphi_{2}\frac{\partial}{\partial\zeta}\left(\varphi_{2}-b_{j}\psi_{2}\right)\right.
\nonumber \\
\left. +p^{\ast}\,\varphi_{1}\frac{\partial}{\partial\zeta}
\left(\varphi_{1}-b_{j}\psi_{1}\right)\right\}dx',
 \label{eqg}
\end{eqnarray}
where, after differentiating, the integrand is evaluated at
$\zeta=\zeta_{j}$. Equations (\ref{eqa}), (\ref{eqb}),
(\ref{zeta}) and (\ref{eqg}) describe the evolution of the
scattering data.

If $p[u,u^{\ast}]$ is a small perturbation, one can substitute the
unperturbed $N$-soliton Jost solutions $\psi$, $\bar{\psi}$, and
$\varphi$ into the right-hand side of (\ref{eqa}), (\ref{eqb}),
(\ref{zeta}) and (\ref{eqg}). This yields evolution equations for
the scattering data in the lowest approximation of perturbation
theory. This procedure can be iterated to yield higher orders of
perturbation theory. The appearing hierarchy of equations is
applied to an arbitrary number of solitons and, in particular,
describes nontrivial many-soliton effects in the presence of
perturbations. In this paper we restrict ourselves to the case of
one-parametric one-soliton solutions with
$\zeta_{1}=\rho\exp(i\beta_{1})$ and substitute unperturbed
one-soliton Jost solutions (\ref{psi1})--(\ref{psi2}) into the
right-hand side of (\ref{eqa}), (\ref{eqb}), (\ref{zeta}) and
(\ref{eqg}). The resulting equations are the desired set
describing the evolution of the scattering data (both solitonic
and continuous) in the presence of perturbations. Under this,
equations (\ref{zeta}) and (\ref{eqg}) corresponds to the so
called adiabatic approximation, when an unperturbed instantaneous
shape of one soliton with slowly varying parameters $\beta_{1}$
and $b_{1}$ is assumed, while equations (\ref{eqa}) and
(\ref{eqb}) account for radiative effects. Making use of the
relation between the soliton solution (\ref{usual}) and the
corresponding squared Jost solution evaluated at $\beta_{1}$, the
adiabatic equation for $\beta_{1}$ can be simplified to
\begin{equation}
\label{db} \frac{\partial\beta_{1}}{\partial
t}=\frac{i}{4}\int_{-\infty}^{\infty}
(p^{\ast}u_{s}-pu_{s}^{\ast})\,dx,
\end{equation}
where $u_{s}$ is the one-parametric soliton solution
(\ref{usual}). Note, that this equation can also be obtained with
the aid of the integral of motion $I_{0}$.

\begin{figure}
\includegraphics[width=6in]{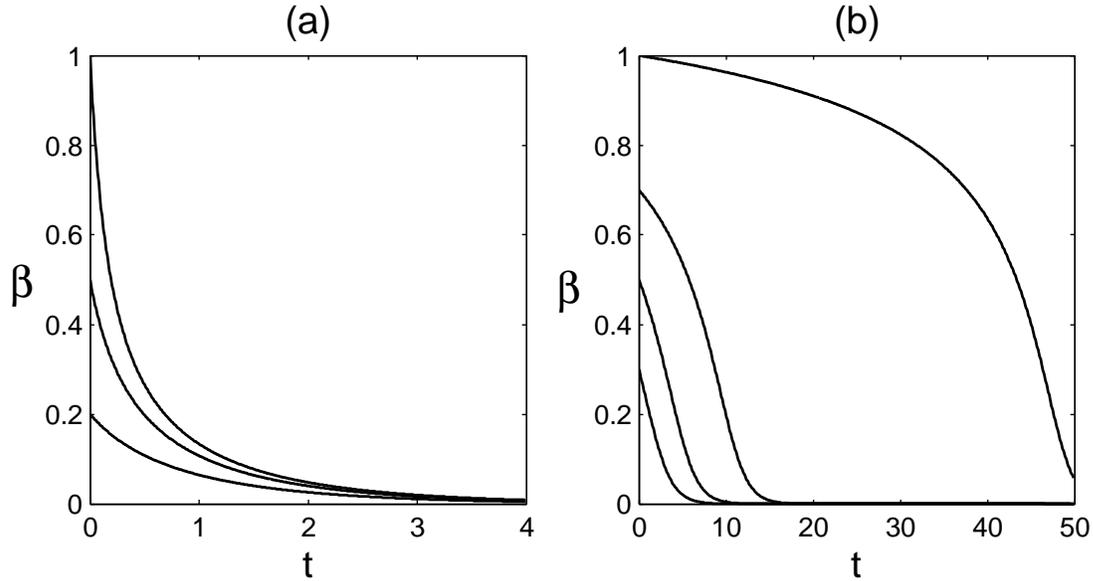}
\caption{\label{fig2} Dependence of the soliton parameter
$\beta(t)$ on time for different initial $\beta(0)$ and  for (a)
bright and (b) dark solitons. }
\end{figure}

\section{Application}
As an example of using of the present perturbation theory, we
consider the case when the perturbation term $p$ in (\ref{mainp})
has the diffusive form
\begin{equation}
\label{p} p=iD\frac{\partial^{2}u}{\partial x^{2}}.
\end{equation}
This form of dissipative perturbation occurs for Alfv\'{e}n
solitons in a plasma when finite electric conductivity (and/or ion
viscosity) of the plasma is taken into account
\cite{Molhus3,Lashkin2}. The conditions (in terms of the plasma
parameters) under which the diffusive term (\ref{p}) can be
considered as a small perturbation are given in
\cite{Molhus3,Lashkin2}. We consider the action of perturbation on
the one-parametric soliton (\ref{usual}) in the adiabatic
approximation. According to this approximation, the parameter
$\beta$ of the soliton (\ref{usual}) is considered as slowly
varying in $t$ but with the unchanged functional shape. Then,
substituting (\ref{p}) into (\ref{db}) and calculating integrals
with $u_{s}$ given by (\ref{usual}), one can obtain
\begin{equation}
\label{evbeta}
 \fl \frac{\partial\beta}{\partial t}=-4D\rho^{2}\sin\beta
[\epsilon\sin\beta
(\cos^{2}\beta-3)(\pi-2\epsilon\beta)+2\cos\beta
(3-2\cos^{2}\beta)]
\end{equation}
Numerically found solutions of (\ref{evbeta}) for different
initial values of $\beta$ are shown in figures ~\ref{fig2}(a) and
~\ref{fig2}(b) for bright ($\epsilon=-1$) and dark ($\epsilon=1$)
solitons respectively. For sufficiently small initial $\beta(0)\ll
1$, from (\ref{evbeta}) one can get a simple estimate
$\beta(t)=\beta(0)\exp(-8D\rho^{2} t)$ both for bright and dark
solitons, so that their amplitudes and velocities decrease with
time. The situation, however, changes dramatically when $\beta(0)$
is not too small. Under this, the behaviour of bright and dark
solitons is essentially different. First of all, as one can see in
figure~\ref{fig2}, dark solitons turn out to be much more robust.
Next, if the initial $\beta$ exceeds the critical value
$\beta_{cr}=\pi/6$ (see figure ~\ref{fig1}), then the amplitude
$A_{d}$ (with respect to the background) of the dark soliton first
increases with time, reaches a maximum for $\beta_{cr}=\pi/6$,
where $A_{d}=\rho$, and finally decreases.

\section{Conclusion}

We have presented a simple approach for finding $N$-soliton
solution and the corresponding Jost solutions of the DNLSE with
NVBC. It is important that the exact solutions can be obtained
without explicit determining of the phase factor. The found one-
and two-soliton solutions perfectly coincide with those obtained
in \cite{Chen1,Chen3}, but, unlike \cite{Chen1,Chen3}, our method
allows to get solutions describing collisions between breathers,
as well as collisions between pure bright/dark solitons and
breathers.

We have also developed a perturbation theory based on the IST for
perturbed DNLSE solitons. This approach fully uses the natural
separation of the discrete and continuous degrees of freedom of
the unperturbed DNLSE with NVBC. We have derived evolution
equations for the scattering data (both solitonic and continuous)
in the presence of perturbations. As an application of the
developed theory, we considered (in the adiabatic approximation)
the action of the diffusive-type perturbation on a single
bright/dark soliton.

\end{document}